\documentclass[prb,preprint,floats,aps]{revtex4}

\usepackage{bm}  
\usepackage{textcomp}
\usepackage{dcolumn}
\usepackage{graphicx}  
\usepackage{amsfonts}  
\usepackage{amsmath}  
\usepackage{amssymb}

\begin{document}
\preprint{}
\title{Plasmonic Excitations in Tight-Binding Nanostructures}

\author{Rodrigo A. Muniz}
\email{rmuniz@usc.edu}
\affiliation{Department of Physics and Astronomy, University of Southern 
California, Los Angeles, CA 90089-0484} 
\author{Stephan Haas}
\affiliation{Department of Physics and Astronomy, University of Southern 
California, Los Angeles, CA 90089-0484} 
\author{A.F.J. Levi}
\affiliation{Department of Electrical Engineering, University of Southern
California, Los Angeles, CA 90089-0082}
\author{Ilya Grigorenko}
\affiliation{Theoretical Division T-11, Center for Nonlinear Studies, 
Center for Integrated Nanotechnologies, Los Alamos National Laboratory, 
Los Alamos, New Mexico 87545, USA}

\date{\today}

\begin{abstract} 
We explore the collective electromagnetic response in 
atomic clusters of
various sizes and geometries. Our aim is to understand, and hence to
control, their dielectric response, based on a fully quantum-mechanical 
description which captures accurately their relevant collective modes.
The electronic energy levels and wave functions, calculated within the 
tight-binding model, are used to determine the non-local dielectric
response function. It is found that the system shape, the electron filling
and the driving frequency of the external electric field strongly control 
the resonance properties of the collective excitations in the frequency and 
spatial domains. Furthermore, 
it is shown that one can design spatially localized 
collective excitations by properly tailoring the nanostructure geometry. 
\end{abstract}

\maketitle

\section{Introduction}
Recent advances in nanoscience have created a vast number of experimentally
accessible ways to configure atomic and molecular clusters into different 
geometries with strongly varying
physical properties.
Specifically, exquisite control of the shape 
and size of atomic and molecular clusters has made it now possible
to investigate
the collective electromagnetic response of ultra-small metal and 
semiconductor particles.\cite{kresin,ho} 
The aim of this study is to model and examine 
plasmonic excitations in such structures, and thus to gain an 
understanding of the 
quantum-to-classical 
crossover of collective modes with increasing cluster size. 
There is obvious technological relevance to tunable collective modes in
nanostructures. For example,
surface plasmon resonances in metallic nanospheres and
films have been found to be highly sensitive to nearby microscopic objects, 
and hence are
currently investigated for potential sensing applications.\cite{homola}
In this context, it is desirable to design customized  
nanostructures with specifically tailored resonance properties,\cite{manoharan} 
and this study is intended to be a first step into this direction.

It is natural to expect that in many cases
the electromagnetic response of nanoclusters 
is considerably different from the bulk. In particular for very 
small clusters, the
quantum properties of electrons confined in the structure need to be
taken into account. 
Moreover, unlike in the bulk, the coupling between 
single-particle excitations
and collective modes can be very strongly affected by its 
system parameters.
This exponential sensitivity
opens up excellent opportunities to optimize the dielectric 
response via tuning 
the cluster geometry and its electron filling. 
For example, by proper arrangement of atoms on a surface one can   
design nanostructures with controllable resonances in the 
near-infrared or visible frequency range.\cite{ho} 
A possible application of such nanostructures 
is the creation of metamaterials with negative refractive index 
at a given frequency.
Furthermore, since geometry optimization of bulk resonators has 
demonstrated minimization of losses in metamaterials 
\cite{shalaev}, it is also interesting to investigate the effect of the 
nanostructure shape on the loss function at a given resonance frequency.

To approach this problem, in this study we
investigate the formation of resonances in generic
 systems of finite
conducting clusters, and examine
how their frequency and spatial dielectric response  
depends on the system size and geometry. 
In particular, 
the non-locality of the dielectric response function in these 
structures 
is important, and will therefore be properly accounted for. 
A similar analysis for the case of small 
{\it metallic} nanostructures was performed recently, using 
an effective mass approximation.\cite{Ilya} 
Here we focus on the opposite limit, namely 
we assume that electrons in the cluster
can be effectively described using  
a tight-binding model.\cite{tb}
Because of the more localized nature of the electronic wave functions 
in this model the overall magnitude of the collective modes are expected
to be strongly suppressed as compared to metallic clusters.

This paper is organized in the following way. In the next section we 
introduce the model and method. For a more extended discussion,
the reader is referred to Ref. \cite{Ilya}. Results for the induced
energy as a function of the driving frequency of an externally applied 
electric
field and the corresponding spatial modulations of the charge density
distribution function
are discussed in the following section. Finally, a discussion of possible 
extentions and applications is given in the conclusion section.

\section{Model}
The interaction of electromagnetic radiation with nanoscale conducting 
clusters is 
conventionally described by semi-classical Mie theory.\cite{mie} 
This is a local, continuum-field model which uses empirical values of the 
linear optical response of the corresponding bulk material, and has been 
applied in nanoparticles to describe plasmon resonances. \cite{mie1} 
However, such a semi-empirical continuum description breaks down beyond a 
certain
degree of roughness introduced by atomic length scales, and thus cannot 
be used to describe ultra-small systems.
In addition, near-field applications, such as surface-enhanced Raman 
scattering, \cite{sers} are most naturally described using a real-space 
theory which includes the non-local electronic response of inhomogeneous 
structures.
Therefore, we will use a recently developed self-consistent and fully 
quantum-mechanical model which fully accounts for the non-locality of the 
dielectric response function.\cite{Ilya}

Specifically, to identify the plasmonic modes in small clusters we 
calculate the total induced energy due to an applied external electric 
field with driving frequency 
$\omega$, and scan for the resonance peaks. 
The induced energy is determined within the non-local linear response 
approximation.

To keep the computational complexity of this procedure at a 
minimum, we use a one-band tight-binding model to obtain the electronic 
energy levels $E_i$ and wave functions $\psi_i({\bf r})$ as a linear 
combination of of s-orbitals
\begin{equation} 
\psi_i({\bf r}) = \sum_{i,j} \alpha_{ij} \varphi({\bf r - R_j}),
\end{equation} 
where $\varphi({\bf r - R_j})$ is the wave function of a s-orbital around 
an atom localized at position ${\bf R_j}$ and $\alpha_{ij}$ are the 
coeficients of the eigenvector (with energy $E_i$) of the Hamiltonian, 
which has the matrix elements 
\begin{equation} 
\langle \varphi({\bf r - R_i}) | H | \varphi({\bf r - R_j}) \rangle  = \left\{
\begin{array}{lr} 
\mu & i=j \\ -t & i,j \; n.n. \\ 0 & otherwise
\end{array}
\right.
\end{equation} 
Here $t$ is the tight-binding hopping parameter. 
The Hamiltonian matrix is diagonalized using the Householder method to 
first obtain a tridiagonal matrix and then a QL algorithm for the final 
eigenvectors and eigenvalues. \cite{NumRec}

Once the electronic wave functions have been obtained,
it is possible to calculate the dielectric susceptibility $\chi({\bf r}, 
{\bf r^\prime}, \omega)$ via
\begin{equation} 
\chi({\bf r}, {\bf r^{\prime}}, \omega) = \sum_{i,j} 
\frac{f({E}_{i})-f({E}_{j})}{{E}_{i}-{E}_{j}-\omega-i\gamma} 
{\psi}_{i}^{*}({\bf r}) {\psi}_{i}({\bf r^{\prime}}) {\psi}_{j}^{*}({\bf 
r^{\prime}}) {\psi}_{j}({\bf r}).  
\end{equation} 
The induced charge density distribution function is then obtained by 
\begin{equation} 
{\rho}_{ind}({\bf r},\omega) = \int \chi ({\bf r}, {\bf r^{\prime}}, \omega) 
( {\phi}_{ind}({\bf r^{\prime}},\omega) + {\phi}_{ext}({\bf 
r^{\prime}},\omega) ) 
d{\bf r^{\prime}},  
\end{equation} 
where in turn the induced potential is given by  
\begin{equation} 
\label{eqn:potential}
{\phi}_{ind}({\bf r},\omega) = \int \frac{{\rho}_{ind}({\bf 
r^{\prime}},\omega)} {|{\bf r}- {\bf r^{\prime}} |} d{\bf r^{\prime}} . 
\end{equation} 
We avoid the large memory requirement to store $\chi({\bf r},{\bf 
r^{\prime}},\omega)$ by calculating the induced charge density 
distribution iteratively via 
\begin{equation} 
\label{eqn:charge}
{\rho}_{ind}({\bf r},\omega) = \sum_{i,j} \frac 
{f({E}_{i})-f({E}_{j})}{{E}_{i}-{E}_{j}-\omega-i\gamma} 
{\psi}_{i}^{*}({\bf r}) {\psi}_{j}({\bf r}) \int {\psi}_{i}({\bf r^{\prime}}) 
{\phi}_{tot}({\bf r^{\prime}},\omega) {\psi}_{j}^{*}({\bf r^{\prime}}) 
d{\bf r^{\prime}},  
\end{equation} 
with ${\phi}_{tot}({\bf r^{\prime}},\omega)={\phi}_{ind}({\bf 
r^{\prime}},\omega)+{\phi}_{ext}({\bf r^{\prime}},\omega)$. 
The integrals are evaluated using a $4^{th}$ order formula obtained from 
a combination of Simpson's Rule and Simpson's 3/8 Rule.  
Eqs. \ref{eqn:potential} and \ref{eqn:charge} are solved 
self-consistently by iterating ${\phi}_{ind}({\bf r},\omega)$ and 
${\rho}_{ind}({\bf r},\omega)$. 
This procedure typically converges in 3-8 steps when starting with 
${\phi}_{ind}({\bf r},\omega)=0$, depending on the proximity to a 
resonance and on the value of the damping constant $\gamma$, which 
throughout this 
paper is chosen as $\gamma = 0.08 t$. 
A much better performance can be achieved when the initial 
${\phi}_{ind}({\bf r},\omega)$ is taken as the solution of a previously 
solved nearby frequency. 
Upon its convergence, the frequency and spatial dependence of the induced 
electric field and the induced energy are obtained using 
\begin{equation} 
{\bf E}_{ind}({\bf r},\omega) = - \nabla {\phi}_{ind}({\bf r},\omega)
\end{equation}  
and
\begin{equation} 
U_{ind}(\omega) = \int {|{\bf E}_{ind}({\bf r},\omega)|}^2 d{\bf r}
\end{equation}  
The observed resonances in the induced energy and charge density
distribution
at certain driving frequencies of the applied electric field correspond 
to collective modes of the cluster. 
In the following, the local induced charge density distribution
is used for analyzing the characteristic 
spatial modulation of a given plasmonic resonance.

\section{Results}

\begin{figure}[h]
\includegraphics[width=8cm]{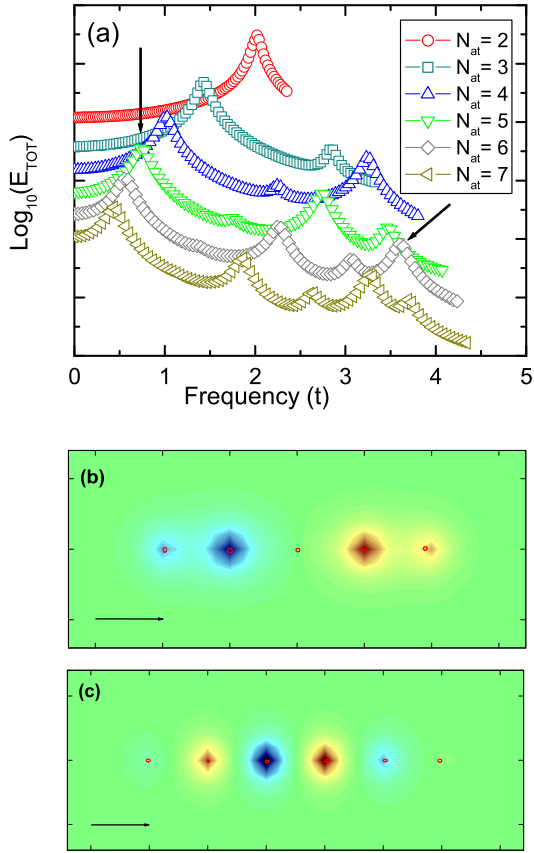}
\caption{\label{fig:Longitudinal} Longitudinal modes in atomic chains.
(a) Decimal logarithm of the total induced energy (artificially offset) as 
a function of the frequency of an external electric field which
is applied along the 
direction of the chain. The resonance peaks correspond to different 
modes. (b) Induced charge density distribution for the lowest energy 
mode at $\omega$ = 0.73$t$ in the 5-atom chain. (c) Induced charge density 
distribution for the highest-energy mode at $\omega$ = 3.56$t$ in the 
6-atom chain.}
\end{figure} 

Let us first focus on the dielectric response
function in linear chains of atoms,
with the intent to identify the basic features of their collective 
excitations. 
The frequency dependence
of the induced energy in such systems, exposed to a driving electric field 
along the chain direction, is shown in Fig.\ref{fig:Longitudinal}(a). 
It exhibits a series of resonances, which increase in number for chains 
with increasing length. 
As observed in the spatial charge density distribution, e.g. shown for the 
5-atom chain in Fig.\ref{fig:Longitudinal}(b), the lowest peak corresponds
to a 
dipole resonance. 
When increasing the system size, the dipole peak moves to lower 
frequencies, which is the expected finite-size scaling behavior. 
The resonances at higher frequency correspond to higher harmonic charge 
density distributions. 
For example, in Fig.\ref{fig:Longitudinal}(c), we show the charge density 
distribution corresponding to the highest frequency resonance of the 
6-atom chain. 
In contrast to the dipole resonance, these modes show a rapidly 
oscillating charge density distribution, and thus have the potential to provide 
spatial localization of collective excitations
in more sophisticated structures.  
While an extension to much larger chains is numerically prohibitive within
the current method, the finite-size scaling of the 
observed dielectric response of these clusters is consistent with the 
1D bulk expectation of a dominant low-energy plasmon mode, coexisting
with a particle-hole continuum of much weaker spectral intensity.

\begin{figure}[h]
\includegraphics[width=8cm]{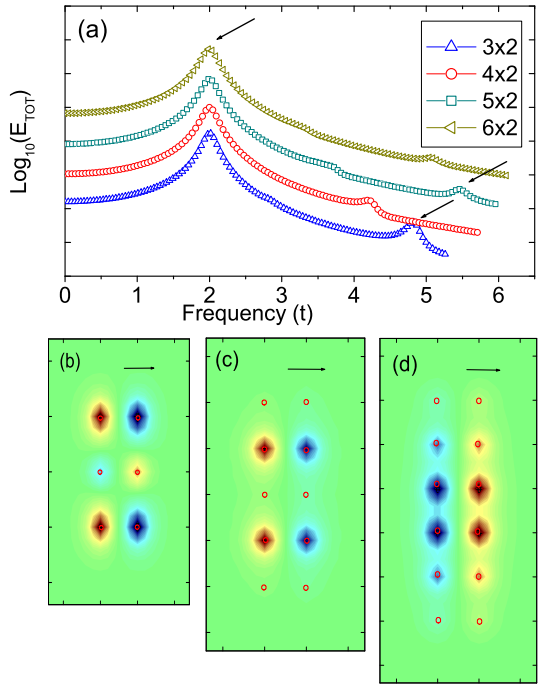}
\caption{\label{fig:Transverse} Transverse Modes in coupled chain structures.
(a) Logarithm of the total induced energy (artificially offset) as a 
function of the frequency of an external electric field applied 
transversely to the chain. The low energy mode is central, the analog of 
a bulk plasmon, and the high energy mode is located at the surface, the 
analog of a surface plasmon. (b) Induced charge density
distribution for the mode 
at $\omega = 4.91t$ in the 3-atom double chain. (c) Induced charge density 
distribution for $\omega = 5.41t$ in the 5-atom double chain. (d) Induced 
charge density distribution for $\omega = 1.93t$ in the 6-atom double chain. }
\end{figure} 

In order to study the transverse collective modes we apply an external 
electric field perpendicular to ladder structures made of 
coupled linear chains of atoms.\cite{footnote} 
Fig.\ref{fig:Transverse}(a) shows that for every chain size there are two 
resonance peaks for the total induced energy, the higher energy is an end 
mode, as shown in Figs.\ref{fig:Transverse}(b) and (c) for the 3 and 5-atom 
double chains respectively, whereas the lower energy peak corresponds to a 
central mode, as displayed in Fig.\ref{fig:Transverse}(d) for the 6-atom 
double chain. 
It is also confirmed that as the length of the chain is increased, the 
central mode gets stronger relatively to the end mode, which is the 
expected behavior for  bulk vs. surface excitations.  
These results are in agreement with the findings of Ref.\cite{Yan}.

\begin{figure}[h]
\includegraphics[width=8cm]{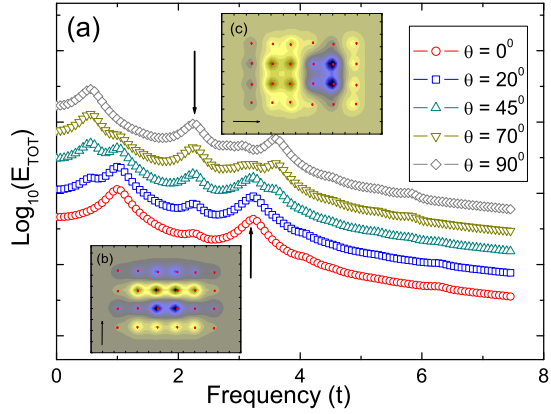}
\caption{\label{fig:Angle} Dependence on the direction of the applied 
electric field. 
(a) Logarithm of the total induced energy (artificially offset) as a 
function of frequency of external electric fields applied to a 
4$\times$6-rectangle at different incident
angles. $\theta = 0$\textdegree when the field 
is parallel to the 4-atoms edge, and $\theta = 90$\textdegree when it is 
parallel to the 6-atoms edge. (b) Induced charge distribution for $\theta 
= 0$\textdegree and $\omega = 3.15t$. (c) Induced charge density
distribution for 
$\theta = 90$\textdegree and $\omega = 2.15t$.}
\end{figure}

Let us next examine what happens when the direction of the 
external electric field is varied. 
Fig.\ref{fig:Angle} shows the dielectric
response of a 4$\times$6-atom rectangular 
structure for different angles incidence directions of the applied field. 
When the field is parallel to one of the edges ($\theta = 0$\textdegree 
or $\theta = 90$\textdegree), the response is essentially that of a 
single chain with the same length, shown in 
Fig.\ref{fig:Longitudinal}(a). 
Also the induced spatial charge density
modulations are analogous to those of the 
correspondent linear chain, which can be seen in Figs. \ref{fig:Angle}(b) 
and (c). 
At intermediate angles the response is a superposition of the two above cases, 
changing gradually from one extremum to the other as the angle is 
changed. 
Notice for instance that as the angle increases, the peak at the same 
frequency of the 4-atoms dipole resonance diminishes, while simultaneously 
another resonance 
is formed at the frequency of the dipole mode of a 6-atoms chain when the 
angle is tuned from $\theta = 0$\textdegree \ to $\theta = 90$\textdegree. 
For $\theta = 0$\textdegree \ there is only the peak at the frequency of 
the 4-atom chain dipole resonance, whereas
for $\theta = 90$\textdegree \ only the 
dipole peak corresponding to the 6-atoms dipole frequency is present. 
The superposition of the responses from each direction is a consequence 
of the linear response approximation employed, since the response is a 
linear combination of those obtained from each direction component of the 
external field.

\begin{figure}[h]
\includegraphics[width=8cm]{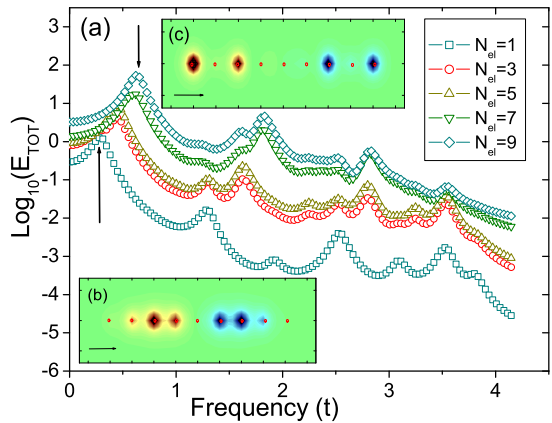}
\caption{\label{fig:Filling} Variation of the number of electrons.
(a) Logarithm of the total induced energy as a function of the external 
electric field frequency. The number of electrons $N_{el}$ in a 9-atom 
chain is varied. (b) Induced charge density distribution for $N_{el} = 1$ at 
$\omega = 0.36t$. (c) Induced charge density distribution for $N_{el} = 9$ at 
$\omega = 0.53t$.}
\end{figure} 

Next, let us analyze the dependence of the resonance modes on the number of 
electrons in the cluster. 
Fig.\ref{fig:Filling}(a) shows significant changes in the response of a 
9-atom chain with the external field applied along its direction.
In particular, 
it is observed that the response is stronger when there are more electrons in 
the sample, a quite obvious fact since there are more particles 
contributing to the collective response. 
Moreover the resonance frequencies of lower modes increase with the 
number of electrons, which can be understood as
a consequence of the one-dimensional tight-binding density 
of states being smallest at the center of the band. 
Hence the energy levels 
around the Fermi energy are more sparse in the finite system,
and therefore the excitations require 
larger frequencies at half-filling.
The same does not hold for higher frequency modes since these 
correspond to transitions between the lowest and highest levels for any 
number of electrons in the sample. Therefore these modes have the same 
frequency, independent of the electronic filling.
Higher filling also allows the induced charge density to concentrate closer to 
the boundaries of the structure, as a comparison between 
Figs.\ref{fig:Filling}(b) and (c) demonstrates.   
Fig.\ref{fig:Filling}(b) shows that a 9-atoms chain with $N_{el} = 1$ 
electron has its induced charge density
localized around the center of the chain. 
In contrast, Fig.\ref{fig:Filling}(c) displays the induced charge
density localized at 
the boundaries of the same structure with $N_{el} = 9$. 
This concentration closer to the surface happens because higher energy 
states have a stronger charge density modulation than the lower energy ones. 
Therefore the induced charge density
is more localized for higher fillings, because 
at low fillings the excitations responsible for the induced charge density are 
between the more homogeneous lower energy levels.
This can be interpreted as
a finite-size rendition of the fact that by increasing the 
electronic filling one obtains the classical response with all the induced 
charge density on the surface of the object.

\begin{figure}[h]
\includegraphics[width=8cm]{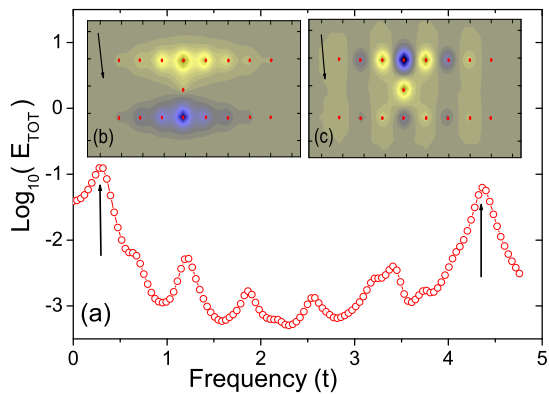}
\caption{\label{fig:Hshape1} Connect between two chains, $N_{el}$ = 1.
(a) Logarithm of the total induced energy as a function of the frequency 
of an external electric field applied to two 8-atom chains with an extra 
atom connecting them at the center. (b) Induced charge density
distribution for 
$\omega$ = 2.28$t$. (c) Induced charge density
distribution for $\omega$ = 4.36$t$.}
\end{figure} 

Access to high energy states is very important for achieving spatial 
localization of the induced charge density, as the next example shows. 
In order to find a structure with spatially localized plasmons we 
consider two parallel 8-atoms chains connected to each other by an 
extra atom at the center.
When an external electric field is applied transversely to the chains, 
the electrons are stimulated to hop between them, but this is only 
realizable through the connect, therefore the plasmonic excitation is 
sharply localized around it. 
Fig.\ref{fig:Hshape1}(a) shows the response of two 8-atoms chains, 
Fig.\ref{fig:Hshape1}(b) and (c) show respectively the induced charge 
density for the dipole and the highest modes.
It is seen that the lowest frequency mode has a much longer spread along 
the chains, whereas the high frequency plasmon is sharper, since it 
corresponds to excitations to the highest energy state that has a large 
charge ondulation as it was pointed out before. 

\begin{figure}[h]
\includegraphics[width=8cm]{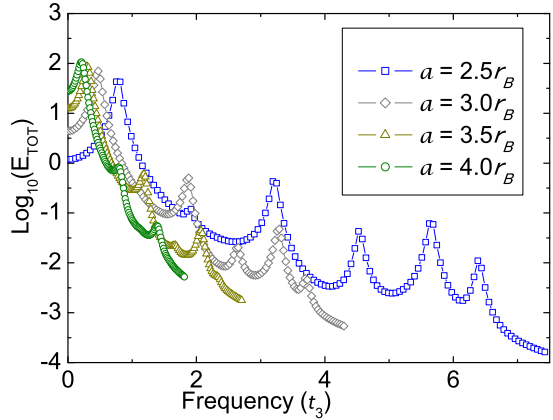}
\caption{\label{fig:Spacing} Variation of the distance between neighbor atoms $a$. Logarithm of the total induced energy as a function of the frequency of an external electric field applied to 7-atom chains with different spacing between atoms $a$ in units of the Bohr radius $r_{B}$. The frequency unit is $t_{3}$ the tight-binding hopping parameter for $a$ = 3$r_{B}$.}
\end{figure}

Let us finish by analyzing the dependence of the various dielectric response modes on the distance between atoms in the chain.
The dipole moment of the chain is proportional to its length, and consequently also proportional to the distance between neighboring atoms. 
Hence one can expect that the strength of the dielectric 
response is also proportional to the atomic spacing. 
However, higher frequency modes require a larger mobility of electrons since they need to move quickly along the chain in order to produce the charge oscillations of the mode. In other words, the charge-density waves can only have frequencies at most equal to the wavering of the charge carriers.
Hence the high frequency modes are suppressed for systems where electrons lack the mobility necessary for the execution of their oscillation.  
In the tight-binding model, the electronic mobility stems from the tunneling of electrons from one atom to another. 
The tunneling rate depends on the overlap of the atomic orbitals located in different sites, and moreover this overlap decays exponentially with the spacing between atoms. 
Therefore the high frequency modes are weaker for long spaced chains, while the slow modes get stronger.
This fact is demonstrated in Fig.\ref{fig:Spacing} where the response of 7-atom chains with different atomic spacings are shown. 
The tight-binding hopping parameter $t$ changes with the atomic spacing $a$,
and here we considered a generic\cite{harrison}
power-law dependence $t \sim a^{-3}$.
The figure clearly shows that the strength of the slowest mode is reduced for shorter spacings. The opposite is true for the two fastest modes, while the intermediate modes have nearly no change.

\section{Conclusion}
In conclusion, we have analyzed the evolution of plasmonic resonances in 
small clusters as a function of the system
shape, applied external fields, electron filling and atomic separation.
Using a fully 
quantum-mechanical, non-local response theory, we observe     
that longitudinal and transverse modes are very sensitive to these system
parameters. This is reflected in their frequency, oscillator strength and 
the spatial modulation of the induced charge density. Specifically,
we identify bulk and surface plasmonic 
excitations which can be controlled in amplitude and frequency by the 
cluster size. Furthermore, we observe a non-trivial filling dependence, 
which critically depends on the electronic level spacing in a given 
structure. We also find that changes in atomic spacings strongly affect 
the electron mobility in these structures, with a
very different impact on low-energy
vs. high-energy modes. And we see that changing
the position of a single atom 
in a nanostructure can completely alter its collective dielectric response.
This strong sensitivity to small changes is the key to controlling
the modes of ultra-small structures, and it can thus become the gateway to a 
new generation of quantum devices which effectively utilize quantum physics
for new functionalities.

\begin{acknowledgments}
We would like to thank Gene Bickers, Richard Thompson, Vitaly Kresin,
Aiichiro Nakano amd Yung-Ching Liang for useful conversations, and we
acknowledge financial support by the Department of Energy, grant number
DE-FG02-06ER46319.
The numerical computations were carried out on the University of Southern 
California high-performance computer cluster.
\end{acknowledgments}


\begin{thebibliography}{200}
\bibitem{footnote} Within the current approach, at least two coupled chains
are necessary to visualize charge redistributions along the transverse 
direction, since charge fluctuations within the orbitals are not accounted for.
\end{thebibliography}
\end{document}